**Magnetic Topological Semimetal Phase with Electronic Correlation Enhancement in SmSbTe**

Krishna Pandey, Debashis Mondal, John William Villanova, Joseph Roll, Rabindra Basnet, Aaron Wegner, Gokul Acharya, Md Rafique Un Nabi, Barun Ghosh, Jun Fujii, Jian Wang, Bo Da, Amit Agarwal, Ivana Vobornik, Antonio Politano*, Salvador Barraza-Lopez*, Jin Hu*

**Abstract**

The ZrSiS family of compounds hosts various exotic quantum phenomena due to the presence of both topological nonsymmorphic Dirac fermions and nodal-line fermions. In this material family, the LnSbTe (Ln= lanthanide) compounds are particularly interesting owing to the intrinsic magnetism from magnetic Ln which leads to new properties and quantum states. In this work, the authors focus on the previously unexplored compound SmSbTe. The studies reveal a rare combination of a few functional properties in this material, including antiferromagnetism with possible magnetic frustration, electron correlation enhancement, and Dirac nodal-line fermions. These properties enable SmSbTe as a unique platform to explore exotic quantum phenomena and advanced functionalities arising from the interplay between magnetism, topology, and electronic correlations.

# 1. Introduction

Topological materials are a promising platform for a wide range of next-generation technologies.[1–16] In topological Dirac or Weyl semimetals, symmetry-protected Dirac or Weyl nodes host relativistic fermions with low-energy excitations that can be described by Dirac or Weyl equations, respectively.[4,5] In addition, topological nodal-line semimetals exhibiting linear band crossings along a 1D loop have also been discovered.[7] Recently, there is a rapidly growing interest in the ZrSiS topological nodal-line semimetal family, represented by the chemical formula WHM (W=Zr/Hf/rare earth, H= Si/Ge/Sn/Sb, M=O/S/Se/Te).[17–30] These materials crystalize in a PbFCl-type structure comprised of 2D square nets of H atoms sandwiched by W-M layers. Two types of Dirac states have been discovered in this material family thus far: (i) a gapless Dirac point state protected by nonsymmorphic symmetry, and (ii) a Dirac nodal-line generated by the glide mirror symmetry, which is gapped by spin-orbit coupling (SOC).[17–20,31,32] Exotic properties such as electronic correlation enhancement[33] owing to the nodal-line structure[34] and a new surface state due to broken symmetry at surfaces[35] which is robust[36] against surface oxidation[36,37] have been reported. In addition to those non-magnetic materials, the spin degree of freedom can be activated in WHM compounds by making W a magnetic lanthanide (Ln) element, H = Sb, and M = Te. Such LnSbTe materials provide a platform to further investigate exotic quantum states.[38–43] New phenomena such as Kondo effects, charge density waves (CDW), and correlation enhancement have been discovered in various LnSbTe compounds.[39,43,44] Those quantum phenomena, together with the non-trivial topology generated by the Sb layer in LnSbTe,[25,26,41,45] offer great opportunities for exploring topological physics and for developing topological quantum material-based devices.

The study of magnetic LnSbTe compounds is in an early stage with only few materials discovered, such as CeSbTe, GdSbTe, NdSbTe, and HoSbTe.[25,26,38–43,45,46] Here, we report the investigation of the previously unexplored SmSbTe member of the family. Our combined magnetization and specific heat measurements revealed antiferromagnetism and signatures of electronic correlation enhancement. These findings, together with the Dirac states revealed by angle resolved photoemission spectroscopy (ARPES) and electronic band structure calculations, suggest that SmSbTe represents an ideal platform for new exotic quantum phenomena arising from the interplay between magnetism, topology, and electronic correlations. The manipulation of these phenomena would further pave a path for quantum material-based functional devices.

## 2. Results and Discussion

SmSbTe single crystals were grown by chemical vapor transport (CVT). The sharp (00L) x-ray diffraction peaks (Figure 1a) reveal the excellent crystallinity of our single crystals. Single crystal x-ray diffraction further reveals that SmSbTe crystalizes in a tetragonal lattice with nonsymmorphic space group P4/nmm, formed by the stacking of a Te-Sm-Sb-Sm-Te slab in which the Sb square nets are sandwiched by Sm-Te layers (Figure 1b). The structural parameters obtained from the refinement are provided in the Supporting Information. Earlier studies on other LnSbTe compounds[38,47,48] revealed possible Te substitution and vacancies in the Sb layer, which reportedly cause CDWs. In our SmSbTe, the composition analysis using energy dispersive x-ray spectroscopy indicates a stoichiometric composition, which is further confirmed by the lack of a CDW order as will be shown later on.

The magnetic order induced by rare earth elements in LnSbTe is predicted to tune the topological electronic states via the coupling between magnetism and band topology,[26] hence enabling functional properties in these materials. The magnetic property characterization of SmSbTe is presented in Figure 2. Similar to other LnSbTe compounds (Ln = Ce, Nd, Gd, and Ho),[26,40–43] SmSbTe displays an antiferromagnetic (AFM) order with a Néel temperature $T_N$ ~ 3.7±0.2 K. The slight variation of 0.2 K is possibly due to tiny variations in compositions of the multiple crystals we measured. The probed $T_N$ is independent of magnetic field strength (Figure 2a, inset) though, a situation in stark contrast with CeSbTe and NdSbTe, where $T_N$ is suppressed upon increasing the magnetic field.[26,39,43] The AFM nature of the magnetic transition can be determined by the absence of irreversibility between zero-field-cooling (ZFC) and field-cooling (FC) measurements in the temperature dependence of the susceptibility $\chi(T)$ (Figure 2a), and by the linear field dependence at low fields and below $T_N$ in isothermal magnetization $M(H)$ (Figure 2b). In addition, $\chi(T)$ is found to follow a modified Curie-Weiss law $\chi_{mol} = \chi_0 + C/(T - \theta)$ above 20 K, where $\chi_0$ is the temperature-independent part of the susceptibility, $C$ is the Curie temperature, and $\theta$ is the Weiss temperature. The negative $\theta$ ~ -31.7 K extracted from the fit (Figure 2a) implies a strong AFM exchange interaction among Sm moments, a behavior analogous to that observed in CeSbTe ($\theta$ = -10 to -30 K)[26,39] and GdSbTe ($\theta$ = -19 K to –24K).[25,40,41] From the obtained $\theta$, the frustration parameter, defined as $f=|\theta|/T_N$, is ≈8, suggesting possible magnetic frustration,[49,50] which is consistent with our specific heat measurement to be discussed later. The fitted Curie constant, $C$, yields an effective moment $\mu_{eff} = \sqrt{3k_B C/N_A}$ = 1.05 $\mu_B$, which is slightly larger than the theoretically expected value of $\mu_{\text{eff}}$ = 0.857 $\mu_B$ for a Sm$^{+3}$ ion with 4$f^5$ configuration. The slightly larger $\mu_{eff}$ may originate from the polarization of conduction electrons, or it may also be due to the reduction of moment density generally seen in frustrated magnetic systems.[49]

As mentioned above, the isothermal magnetization $M$ of SmSbTe is linear in $H$ at low field. With increasing magnetic field, a small deviation from linearity can be seen with out-of-plane fields ($H$//$c$), as shown in the inset of Figure 2b. This implies a possible metamagnetic transition which has also been observed in other magnetic LnSbTe materials such as CeSbTe, NdSbTe, HoSbTe, and GdSbTe.[25,26,39,40,42,43,51] On the other hand, under an in-plane magnetic field ($H$//$ab$) configuration, a linear $M(H)$ is observed for up to 9 T (the highest field of our instrument). Generally, in AFM systems, a large exchange interaction is expected for linear $M(H)$ without polarization up to high field, which appears inconsistent with the low $T_N$ measured in SmSbTe. On the other hand, linear $M(H)$ and low $T_N$ have indeed been observed in other AFM rare earth compounds such as $Sm_2BaPdO_5$ and $GdFe_{0.17}Sn_2$.[49,52] Further theoretical and experimental studies are needed to fully clarify the magnetism in SmSbTe.

Despite of the unambiguous signature of magnetic order in magnetization measurements, the temperature dependence of the resistivity, shown in Figure 2c, does not display a clear feature at $T_N$ (black arrow in Figure 2c). A Hall effect study reveals an electron-type conduction with a carrier density in the order of $10^{21}$ cm$^{-3}$ for SmSbTe (Figure 2d), which is comparable to reported values for CeSbTe [39] and NdSbTe.[43] Similarly, no clear signature is observed in the Hall effect across $T_N$ (~3.7 K). These observations imply that electronic transport may not be strongly coupled with magnetic order. Nevertheless, the non-metallic transport with a logarithmic temperature dependence between 10 and 35 K (Figure 2c) is suggestive of Kondo physics, which has also been proposed in CeSbTe[39] and NdSbTe.[43] Such a possible Kondo effect, to be discussed later, offers opportunities for the realization of new quantum states and for corresponding advanced functional properties.

Specific heat measurements provide further insight into the magnetic properties of SmSbTe. The specific heat divided by temperature $C(T)/T$ data displays a broad peak around 3.2 K (Figure 3a), lower than the magnetic transition temperature probed by magnetic susceptibility in figure 2. This is in sharp contrast with the well-defined specific heat peak at the magnetic phase transition temperature observed in other LnSbTe materials such as NdSbTe,[43] CeSbTe,[26,39] and HoSbTe.[42] Such broad specific heat peak is unlikely to originate from sample quality issues such as inhomogeneity or defects, because it is reproducible in multiple SmSbTe crystals from the same or different growth batches with well characterized compositions, and it does not change after thermal annealing. The profile and position of the broad specific heat peak are essentially field-independent, as shown in the inset of Figure 3a. Such field-insensitivity was also observed in the magnetization measurements shown in inset of Figure 2a. The absence of a clear ordering feature in the specific heat is a common signature of frustrated magnetic materials, where the moments cannot order onto a unique, lowest energy state.[53–55] Such scenario of magnetic frustration agrees with the sizable frustration parameter derived from magnetization measurements mentioned above. Frustrated magnetism is known to provide opportunities to stem abundant exotic phenomena such as skyrmions[56] and quantum spin liquids[57], further enriching the SmSbTe material platform with additional functional properties. Geometric frustration is known to exist in trigonal or hexagonal lattices, while the tetragonal structure of SmSbTe and the lack of strong field dependence in specific heat measurements (Figure 3, inset) appear inconsistent with known geometrically frustrated systems.[58,59] Therefore, frustrated magnetism, if it exist in tetragonal SmSbTe, could most likely be caused by the competition between the nearest-neighbor and next-nearest-neighbor magnetic exchange couplings, which has been proposed[60] and observed in tetragonal FeSe and CoSe.[50,58] To clarify the origin of the broad specific heat peak in SmSbTe, which is distinct from all other known LnSbTe materials, more experimental and theoretical efforts are needed.

The most intriguing feature in our specific heat measurements is the significantly enhanced Sommerfeld coefficient, which can be obtained by separating the electronic specific heat from the total measured value. The total specific heat, $C_{tot}$, can be expressed by $C_{tot} = C_m + C_{ele} + C_{ph}$, where $C_{ele} = \gamma T$ is the electronic specific heat, $\gamma$ is the Sommerfeld coefficient, $C_m$ and $C_{ph}$ are the magnetic and phonon contributions to the specific heat, respectively. Generally, its phonon contribution can be expressed by $C_{ph} = \beta T^3$ at the low temperature limit where $T \ll$ Debye temperature $\Theta_D$, and thus may be separated by a linear extrapolation of $C/T$ vs $T^2$ or by a polynomial fitting. However, these approaches may not work for SmSbTe due to a pronounced specific heat upturn at around 10 K in $C/T$ (Figure 3a). Therefore, to obtain the precise electronic specific heat, we used the isomorphous non-magnetic LaSbTe as a reference sample, and followed the principle of corresponding states to evaluate the phonon contribution $C_{ph}$, which has been demonstrated to be a justified approach in our previous specific heat studies on NdSbTe[43] and Fe(Te,Se).[61] As shown in Figure 3a, our analysis yields a Sommerfeld coefficient $\gamma$ of 160 mJ/mol, as well as a phonon coefficient of $\beta = 0.465$ mJ/mol K$^4$ which yields a Debye temperature $\Theta_D$ of 232.14 K. With the obtained electronic specific heat $C_{ele} = \gamma T$ and phonon specific heat $C_{ph}$, the magnetic specific heat $C_m$ of SmSbTe can be extracted, from which the magnetic entropy $S_m = \int_0^T \frac{C_m(T)}{T} dT$ is also obtained, as shown in Figure 3b. With increasing temperature, the entropy quickly rises, saturating to 4.33 J/mol K. Assuming a J = 5/2 doublet for $Sm^{3+}$ in the crystal field of SmSbTe, such entropy is 76% of the expected value of Rln2 where R is gas constant, which implies a possible residual magnetic entropy that is widely observed in frustrated systems.[62,63]

The Sommerfeld coefficient, $\gamma$, in LnSbTe is strongly dependent on the choice of the Ln element. $\gamma$ is as small as 0.51 mJ/mol $K^2$ in non-magnetic LaSbTe,[43] and it increases to 7.6 mJ/mol $K^2$ in GdSbTe,[40] and to 10−40 mJ/mol $K^2$ in CeSbTe.[26,39] In addition to $\gamma$ = 160 mJ/mol for SmSbTe as determined in this work, $\gamma$ above 100 mJ/mol $K^2$ has also been observed in NdSbTe (115 mJ/mol $K^2$) [43] and HoSbTe (382.2 mJ/mol $K^2$).[42] The drastic enhancement of $\gamma$ in SmSbTe implies a large effective mass which, as will be shown later, might be associated with the presence of flat Sm 4$f$ bands around the Fermi level ($E_F$). This finding further highlights the importance of electron interactions in SmSbTe. Its coupling with magnetism and topological bands (shown below) further offers a new route of manipulating quantum phases in this material.

In addition to magnetism and electronic correlation enhancement, the existence of topological relativistic fermions revealed by our ARPES experiments further enables SmSbTe as a versatile platform with potential to realize advanced functionalities arising from the interplay between various quantum degrees of freedom. Figure 4a shows the Fermi surface and constant energy contour plots in the paramagnetic (PM) state ($T$ = 78 K) using a photon energy of 30 eV on the (001) plane of the crystal (see Experimental Section for details). A diamond-shaped Fermi surface centered at $\Gamma$ is observed, reminiscent of that seen in topological semimetal ZrSiS.[18,21,28] The constant energy contour plots reveal that no other electronic bands approach the $E_F$ down to the 250 meV binding energy, where an extra four-leaf clover-shaped pattern appears around $\Gamma$. These constant energy contours are well reproduced by our density functional theory (DFT) calculations (Figure 4b), in which the $E_F$ position was adjusted (shifted up) by 0.275 eV in order to reproduce the experimental data, as we will further discuss below.

The electronic band structure measured along the $\bar{M} - \bar{\Gamma} - \bar{M}$ and $\bar{X} - \bar{\Gamma} - \bar{X}$ high symmetry directions (Figure 4c, with its related bulk Brillouin zone shown in Figure 4$f$) confirms that the diamond-shaped Fermi surface originates from the linearly dispersing Dirac-like bands. The linearly dispersing bands crossing $E_F$ are clearly observed along both momentum directions. In particular, Dirac points at 0.275 eV binding energy are observed along $\bar{\Gamma} - \bar{M}$, as indicated by the black arrows. The observed electronic band structure agrees well with DFT calculations on a slab geometry (see Experimental Section) as shown in Figure 4**e**. The band structure calculation of the bulk bands (Figure 5a) further revealed that the observed Dirac bands belong to the Dirac nodal-line generated by $C_{2v}$ symmetry, which are formed mainly by the Sb p-electrons (see Supporting Information). A comparison between the ARPES results and the calculated bands indicates that $E_F$ of SmSbTe lies 0.275 eV above the Dirac crossings at the $\bar{\Gamma} - \bar{M}$ cut. Similar to other compounds in the ZrSiS-family, the nodal-line crossings in SmSbTe are expected to be slightly (~55 meV) gapped by SOC, as revealed by our bulk band calculations shown in Figure 5a. Nevertheless, gap opening at the crossing points is not observed in our ARPES data. An undetectable SOC gap is also reported in recent ARPES study on LaSbTe, possibly due to the limitation of the instrument resolution.[64] On the other hand, gapped Dirac crossings have been probed in other LnSbTe compounds.[38,44,45] The sizable gap has been proposed to originate from the formation of a CDW order in non-stoichiometric compounds, which enlarges the gap at the Dirac crossings.[38,44] Given that the CDW order occurs in the presence of Te substitution and vacancies in the Sb layer in LnSbTe compounds,[38,41,44,48] the undetectable gap in our ARPES study confirms the nearly stoichiometric composition and absence of structural distortions in our SmSbTe crystals.

Our ARPES observations and DFT calculations (with a shifted $E_F$) indicate that SmSbTe is very similar to ZrSiS, with the Fermi surface of the bulk electronic states formed by the Dirac nodal-line bands only.[17,18,21,28] However, the energy dispersion of the nodal-line states in SmSbTe does not depend on the ARPES photon energy, as shown in Figures 4c**,**d which present the band dispersion measured with photon energies of 30 eV (i.e., $k_z = 0$) and 73 eV (corresponding to 17% of the sixth Brillouin zone), respectively. More band structures measured with various photon energies are shown in the Supporting Information. Such observation implies that the Fermi surface of SmSbTe consists of a 2D sheet extending along $k_z$, which is different from that of ZrSiS. A similar 2D-like Fermi surface sheet for bulk electronic states has also been observed in CeSbTe.[44] Another major difference between SmSbTe and ZrSiS is the location of $E_F$, which lies almost at the Dirac crossing in ZrSiS, but is located 0.275 eV above the nodal-line Dirac crossings in SmSbTe as discussed above. Such a shift of $E_F$ leads to a clean Fermi surface of relativistic fermions without interference from conventional electronic bands in SmSbTe. The shift of $E_F$ has been observed in GdSbTe with non-stoichiometric compositions.[38] In our SmSbTe samples, the composition non-stoichiometry is minimal as discussed above. The origin of such $E_F$ shift is unknown, but possibly attributed to very slight non-stoichiometry composition, or related to the contribution of Sm 4*f* bands which cannot be fully captured in our DFT calculation of the energy band structure in the high temperature PM state.

The calculated electronic band structure for SmSbTe also provides insight into the electronic correlation enhancement probed in specific heat measurements: We performed DFT calculations in the AFM ground state to reveal an AFM ground state that is consistent with the magnetic property characterization described above. Figure 5b shows the calculated band structure for the AFM bulk SmSbTe with SOC, based on an A-type AFM order with out-of-plane Sm moments aligned parallel within each plane but antiparallel between adjacent planes. Such magnetic structure has been adopted in other LnSbTe materials[45] and experimentally confirmed in CeSbTe.[26] Generally, an A-type AFM structure with in-plane FM order is not compatible with magnetic frustration. Nevertheless, there can be other mechanisms leading to frustration in A-type AFM compounds, such as the electron-electron interactions and ligand distortions in $KCuF_3$.[65] Distinct from the PM state in which the Sm 4*f* bands are located far below the energy range shown in Figure 5a, we have observed the formation of 4*f* flat bands near the $E_F$ in the AFM state. As compared with the non-SOC bands (see Figure S5b in the Supporting Information), the strong SOC of *f* electrons drastically splits the *f*-band, leading to the presence of the 4*f* flat bands at the calculated $E_F$ (Figure 5b). As stated above, ARPES observations indicate a shift of $E_F$ in the PM state. Although further ARPES studies are needed to determine the exact location of $E_F$ in the AFM state, the flat band extends onto Dirac nodal-line bands, which opens the possibility of hybridization between the *f*-orbitals and the conduction nodal-line fermions. Such possible hybridization is reminiscent of a mass renormalization via a Kondo mechanism in heavy fermions, which agrees with the mass enhancement probed in specific heat measurements. As mentioned above, our transport measurements on SmSbTe probed signatures of Kondo localization, though the carrier density of ~$10^{21}$ cm$^{-3}$ extracted from Hall effect measurement may to be insufficient to fully screen the localized moments in LnSbTe materials.[44]

In principle, SmSbTe in the high temperature PM state is a topological semimetal with a SOC-gapped nodal-line state similar to that reported on ZrSiS. Further, by opening a band gap at the nodal-line and tunning $E_F$ into the band gap, SmSbTe could become a weak topological insulator consisting of stacked 2D topological insulators.[17,45] One practical approach toward stacked 2D topological insulators is composition engineering, which can induce a CDW gap and tune $E_F$ [38,44]. In the low temperature AFM state, each layer of SmSbTe possess parallel magnetic moments for Sm which breaks time-reversal symmetry. Therefore, the bulk AFM SmSbTe can be viewed as the stacking of 2D Chern insulators, which has been proposed in HoSbTe.[45] The rich topological phases in SmSbTe, coupled with charge and magnetic order offers a new playground to study topological physics.

## 3. Conclusion

We have demonstrated that the newly discovered SmSbTe is a promising material platform with a combination of exotic properties, including 4$f$ magnetism with possible magnetic frustration and Kondo effects, electronic correlation enhancement, and a Fermi surface consisting of Dirac bands only. Such properties convey unique opportunities to study new quantum states and phenomena arising from the interplay among spin, charge, lattice, and topological degrees of freedom. In addition to the apparent coupling between the topological states and magnetism discussed above, the electronic correlation enhancement further provides opportunities to engineer topological electronic states. For example, the electronic correlations have been found to be associated with the Sm 4$f$ bands in the AFM state. Thus, the tuning of magnetism via substitution with other rare earth elements could be an efficient way to control electronic correlations. Furthermore, Kondo physics may also enabled by tuning carrier density, pressure/strain, and high magnetic fields, which are known to tune the competition between Ruderman-Kittel-Kasuya-Yosida and Kondo effects and lead to various quantum states in heavy fermion materials. All of, these knobs may possibly open a route to realize correlated topological quantum phases in SmSbTe.

With a unique coexistence of degrees of diverse quantum freedom and their interplay as mentioned above, we conclude that SmSbTe is a promising platform not only for searching new quantum states and phenomena, but also for exploring technology applications exploiting its tunable topological quantum states.

## 4. Experimental Section

SmSbTe single crystals were grown by a two-step chemical vapor transport (CVT) method. First, constituent elements with stoichiometric ratios were used to grow polycrystals at 700 °C.

Secondly, the finely grinded polycrystals were used as a source material for CVT growth under a temperature gradient of 1000 °C to 900 °C, with Iodine being the transport agent. Square-like crystals with metallic luster can be obtained after 15 days, as shown in the insert of Figure 1a.

The elemental analysis was performed by energy dispersive x-ray spectroscopy. The crystal structure at room temperature was determined by single crystal x-ray diffraction at $298\pm2$ K. Data collection was performed under a nitrogen gas atmosphere at 90 K using a Bruker SMART APEX-II diffractometer equipped with a CCD area detector and graphite-monochromated Mo-$K_\alpha$ radiation ($\lambda$ =0.71073 Å). Data reduction and integration, together with global unit cell refinements were performed using the APEX2 software. Multiscan absorption corrections were applied. The atomistic structure was determined by direct methods and refined by full matrix least-squares methods on $F^2$ using the SHELX package with anisotropic displacement parameters for all atoms. In addition to single crystal x-ray diffraction, powder diffraction was also performed to ensure no impurity phases were present. For powder diffraction, carefully selected single crystals were ground into powder. A comparison between x-ray diffraction patterns obtained from a single crystal, powdered samples, and simulations are shown in the Supporting Information.

Magnetization, specific heat, and electronic transport measurements were performed by a physical properties measurement system (PPMS). The electronic band structure was measured by ARPES at the APE-LE beamline at the Elettra Synchrotron. Reproducible spectra were obtained from multiple SmSbTe crystals. The authos probed the electronic structure by using 26 to 86 eV excitation energy; a comparison with the electronic band structure calculations indicates that the bulk $\Gamma$ is reached at a 30 eV photon energy.

DFT calculations were performed using Vienna ab-initio simulation package (VASP) with projector-augmented wave (PAW) pseudopotentials and the Perdew-Burke-Ernzerhof generalized gradient approximation (PBE-GGA) for the exchange-correlation functional. Calculations were performed with and without SOC for both bulk and slab geometries. For the PM case, the Sm pseudopotential was 4*f*-in-core, while for the AFM case it was 4*f*-in-valence and we employed a Hubbard-correction in Dudarev's approximation for the *f*-electrons (U = 6 eV).[66–71] A k-mesh sampling of 12 × 12 × 6 was used for the self-consistent calculations in the bulk, along with a cutoff energy of 300 eV. The slab geometry consisted of five unit cells in the c direction with a vacuum spacing greater than 20 Å to prevent interaction between periodic copies.

Note added: after the acceptance of this work, the authors became aware of an ARPES study on the same material system.[72]

**Supporting Information**

Supporting Information is available from the Wiley Online Library or from the author.

**Acknowledgements**

J. H. and S. B.-L. acknowledge the support from US Department of Energy, Office of Science, Basic Energy Sciences program under Award No. DE-SC0019467 for the support of crystal growth, property characterizations (magnetic, transport, and thermal), and band structure calculations. J.W.V. acknowledges funding from the US Department of Energy, Office of Basic Sciences (Award DE-SC0016139). Calculations were performed at Cori at NERSC, a DOE facility funded under contract No. DE-AC02-05CH11231. R.B. acknowledges the support from the Chancellor's Innovation and Collaboration Fund at the University of Arkansas. B. D. thanks the support by JSPS KAKENHI Grant Number JP21K14656 and by Grants for Basic Science Research Projects from The Sumitomo Foundation. A. A. and B. G. acknowledge the Science and Engineering Research Board (SERB), Department of Science and Technology (DST) of the government of India for financial support and High-performance computing facility at IIT Kanpur for computational support. This work has been partly performed in the framework of the nanoscience foundry and fine analysis (NFFA-MUR Italy Progetti Internazionali) facility.

**Conflict of Interest**

The authors declare no conflicts of interest.

**Data Availability Statement**

The data supporting to the findings of this study is available from the corresponding author upon reasonable request.

**Figure 1**

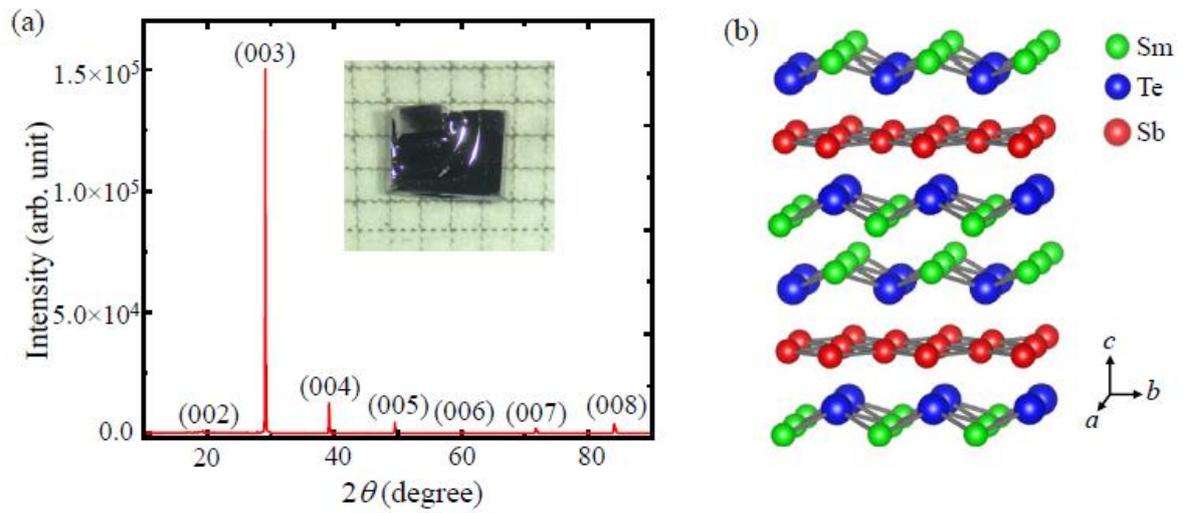

Fig. 1: (a) X-ray diffraction patterns for a SmSbTe single crystal, showing the (00L) reflections. Inset: an image of a single crystal. The square mesh measures 1mm$^2$. (b) Crystal structure of SmSbTe determined by x-ray diffraction. The structural parameters are provided in Supporting Information.

**Figure 2**

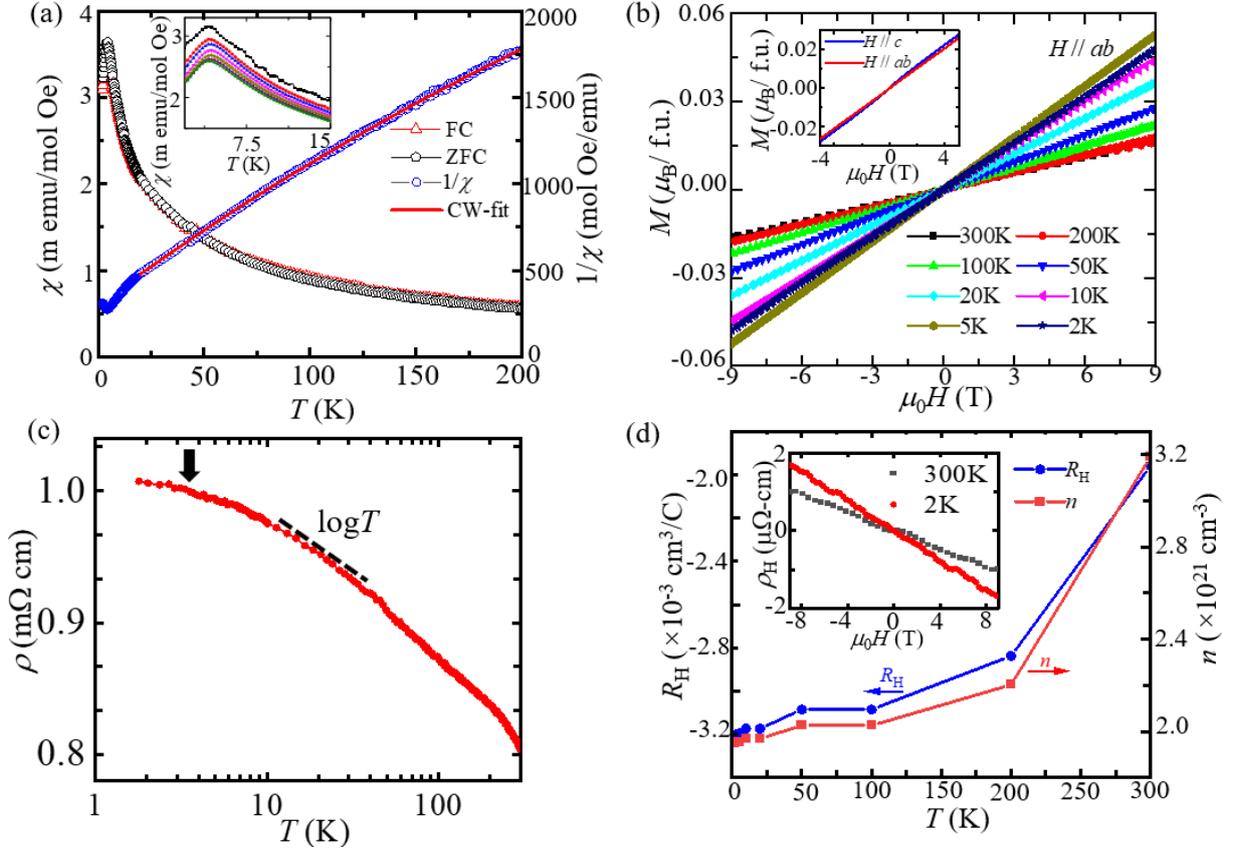

Fig. 2. Magnetization and electronic transport properties of SmSbTe. (a) Temperature dependence of molar susceptibility $\chi$ and inverse susceptibility $1/\chi$ of SmSbTe under an in-plane ($H\|ab$) magnetic field of 1T. The lack of irreversibility between zero field cooling (ZFC) and field cooling (FC) measurements is apparent. The solid, monotonically increasing red line represents the modified Curie-Weiss (CW) fit obtained from $1/\chi$ (see text). Inset: magnetic susceptibility under magnetic fields ranging from 0.1 T (uppermost curve) to 9 T (lowermost curve). (b) Field dependence of magnetization at different temperatures, taken under $H\|ab$. Insets: magnetization for $H\|c$ and $H\|ab$ at T = 2 K. (c) Temperature dependence of the in-plane resistivity of SmSbTe. A logarithmic temperature dependence can be observed 10 and 35 K. The dashed line is a guide to the eye. (d) Temperature dependence of Hall coefficient, $R_H$, and the corresponding carrier density ($1/eR_H$). Inset: field dependence of the Hall resistivity at 2 K and 300 K.

**Figure 3**

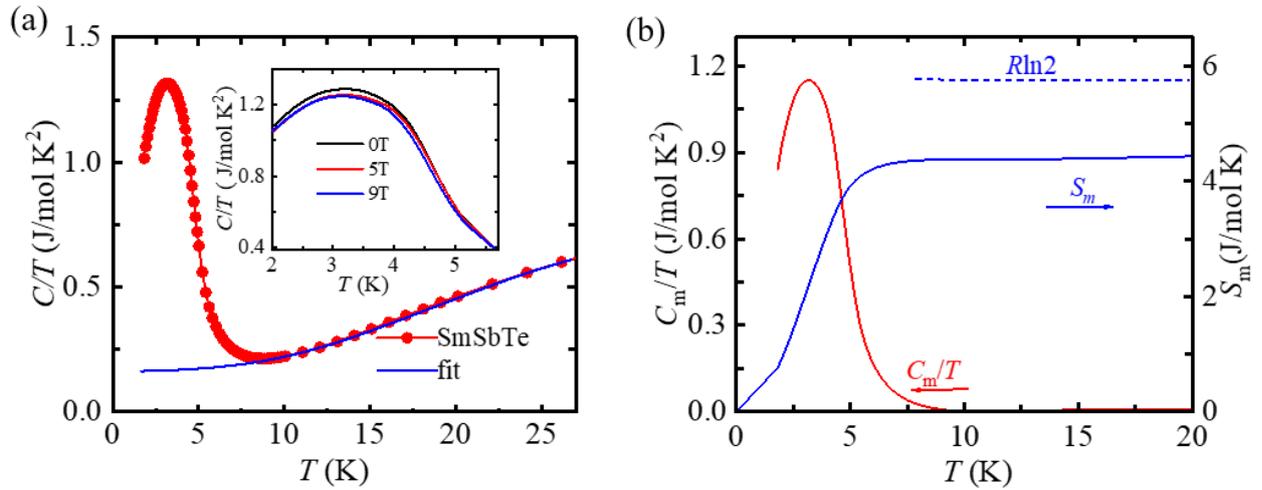

Fig.3 Specific heat of SmSbTe. (a) Temperature dependence of specific heat. The blue solid line represents the fit according to the corresponding state principle using LaSbTe as a reference sample. Data up to 30 K is shown for clarity; data over a larger temerature range is provided in the Supporting Information. Inset: specific heat measured under various magnetic fields increasing from top to bottom. Data for additional fields are provided in the Supporting Information. (b) Magnetic specific heat divided by temperature, $C_m/T$, and the corresponding magnetic entropy, $S_m$.

**Figure 4**

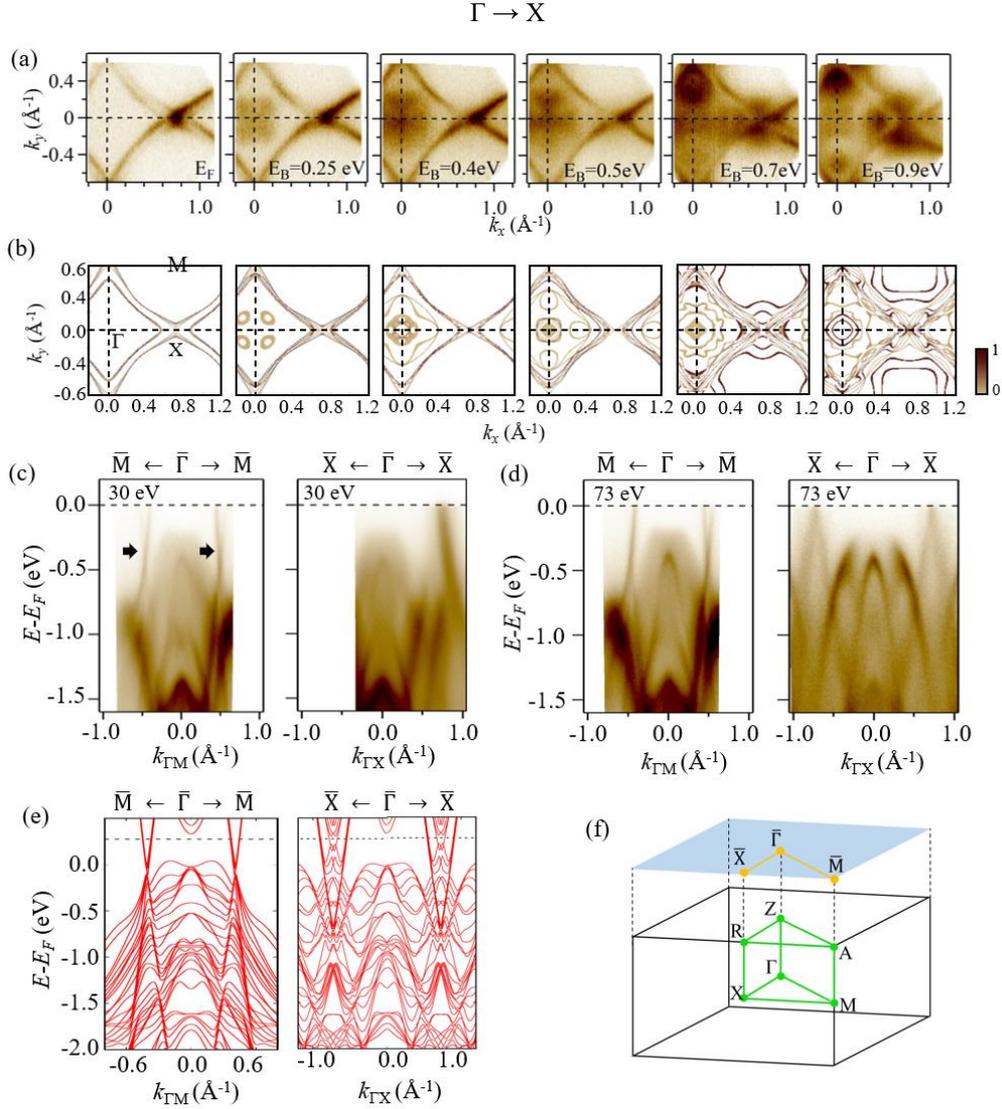

Fig. 4 Electronic structure of SmSbTe in the high temperature PM phase. (a) The Fermi surface and constant energy contours of SmSbTe measured with photon energy of 30 eV at T = 78 K. A four-leave clover feature is seen at (0,0) on the second subplot. (b) Calculated Fermi surface and constant energy contours of SmSbTe with $k_z = 0$ corresponding to the measured spectra in (a). The color scale indicates the surface character, with 0 (light brown) referring to bulk state and 1(dark brown) referring to surface states; see ref [71]. (c) Band structures of SmSbTe measured with a 30 eV photon energy along the $\bar{M} - \bar{\Gamma} - \bar{M}$ and $\bar{X} - \bar{\Gamma} - \bar{X}$ directions. Black arrows indicate the linear band crossing points along the $\bar{M} - \bar{\Gamma} - \bar{M}$ path. (d) Band structure of SmSbTe measured with a 73 eV photon energy. (e) Band structure of five-unit-cell SmSbTe slabs along the $\bar{M} - \bar{\Gamma} - \bar{M}$ and $\bar{X} - \bar{\Gamma} - \bar{X}$ directions with SOC in the PM state. The dashed horizontal line is the $E_F$ experimentally determined by ARPES, which is shifted by 0.275 eV as compared with the calculated $E_F$. (f) Bulk and slab Brillouin zones for SmSbTe.

**Figure 5**

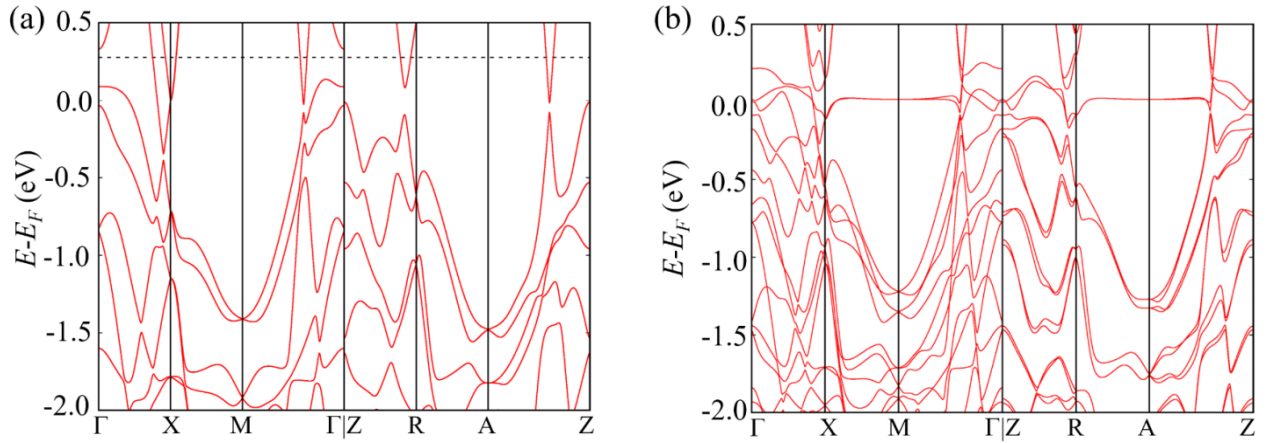

Fig. 5 (a) PM state electronic band structure of SmSbTe with SOC. The dashed lines indicate the shift of the Fermi energy of 0.275 eV from the comparison between the ARPES spectrum (Figure 4) and the calculated band structure in the PM state. (b) AFM state electronic band structure of SmSbTe with SOC. Note that the upper split flat band is beyond the energy range shown in the plot (see Figure S5 in Supporting Information for an AFM band structure without SOC).